\documentclass[aps,12pt]{revtex4}
\usepackage{epsfig}
\usepackage{amsmath}
\usepackage{subfigure}
\usepackage{ulem}
\usepackage{amsfonts}
\usepackage{epsfig,bm}
\usepackage{graphicx}
\usepackage{comment}
\newcommand{\be}{\begin{equation}}
\newcommand{\ee}{\end{equation}}
\newcommand{\ba}{\begin{eqnarray}}
\newcommand{\ea}{\end{eqnarray}}
\newcommand{\nn}{\nonumber}

\begin{document}

\title{\Large{Fermion Fields in BTZ Black Hole Spacetime and Entanglement Entropy }}

\author{Dharm Veer Singh}\email{veerdsingh@gmail.com}
\author{Sanjay Siwach} \email{sksiwach@hotmail.com}

\affiliation{Department of Physics, Centre of Advanced Studies,
Banaras Hindu University
Varanasi- 221 005, India}

\vspace{2.5cm}
\noindent
\begin{abstract}
\begin{center}
ABSTRACT
\end{center}
We study the entanglement entropy of fermion fields in BTZ black hole space-time and calculate pre- factor of the leading and sub-leading terms and logarithmic divergence term of the entropy using the discretized model. The leading term is the standard Bekenstein-Hawking area law and sub-leading term corresponds to first quantum corrections in black hole entropy. We also investigate the corrections to entanglement entropy for massive fermion fields in BTZ space-time. The mass term does not affect the area law. 
\end{abstract}

\maketitle
\newpage

\section{Introduction}

The laws of black hole thermodynamics capture the essential features of macroscopic description of black holes in general theory of relativity. The surface gravity is constant over the horizon and it is related to the temperature of the black hole known as the Hawking temperature. In the context of black hole thermodynamics, the entropy is proportional to area of the event horizon \cite{JD,JD1,JD2,JD3,SWH}. Quantum mechanically black holes emit Hawking radiation and an understanding of this process will lead to the resolution of information loss paradox. If one considers quantum fields in the vicinity of black holes, the area law of entropy receives quantum corrections (logarithmic corrections) \cite{Eisert:2008ur}. The entanglement thermodynamics can also be used to study such corrections. The AdS/CFT correspondence provides a geometric way to calculate the entropy of the black hole known as holographic entanglement entropy \cite{Ryu:2006ef}. Alternatively one can consider the matter fields in the background of black hole \cite{Cadoni:2007vf,Cadoni:2009tk,Cadoni:2010vf} and calculate the entanglement entropy.

The entanglement entropy is the source of the quantum information and it measures the correlation between subsystems separated by the boundary. The entanglement entropy depends upon the geometry of the boundary called the entangling surface. The entanglement entropy is defined by the von Neumann entropy relation ($S_A=-tr_A[\rho_A\ln\rho_A]$), where $\rho_A$ is the reduced density matrix of the system A and is dominated by short range correlations across the entangling surface. These correlations give an area law (Entanglement entropy proportional to the area of entangling surface divided by the cutoff ($\epsilon$)) and the sub leading term in the entanglement entropy contains useful cutoff independent information about the quantum corrections. 

The entanglement entropy approach was first used by Bombelli et al \cite{Bombelli} and Srednicki \cite{Srednicki} for scalar field in spherical systems by considering the entangling surface as would be horizon of a black hole, formed by a collapsing star. Peschel \cite{Peshel} developed a technique to calculate the entanglement entropy of fermions, where the reduced density matrix can be written in terms of fermion correlator. The reduced density matrix in term of correlators obeys the wick theorem and fixes $\rho=c\exp({-\cal H})$, where ${\cal H}$ is a Hamiltonian, quadratic in fields. In this paper, we calculate the density matrix for fermions fields in BTZ space-time and explicitly diagonalize the reduced density matrix in order to estimate the entanglement entropy. 

We also consider the massive fermion fields in BTZ black hole space-time and calculate the entanglement entropy. We expand the entropy in powers proper distance from horizon for different masses and calculate the coefficients of the $\frac{1}{\mu \rho}$ term using the entanglement entropy approach \cite{H2}. The correction to entanglement entropy for massive fields in $(2+1)$ dimension is related to the coefficient of logarithmic term in $(3+1)$ dimensional massless theory \cite{DNK:1994}. The area law contribution of entanglement entropy is not affected by the mass term and the universal quantities depend upon the basic properties of the system (spatial dimension) \cite{Myers,HC3,MPH,Hertzberg:2012mn}. 

This paper is organized as follows. We study the fermion field propagating in BTZ black hole space-time in section (II). In section (III), we calculate the co-efficients appearing in the entropy formula numerically. The contribution of massive fermion fields is studied in section (IV). Finally, we summarize our results and their physical implications.

\section{Fermions in BTZ Black Hole space-time}
The BTZ black hole is a solution of (2+1) dimensional gravity with negative cosmological constant ($\Lambda=-\frac{2}{l^2}$) \cite{MB} and the metric is given by;
\be
{ds}^2=-N^2(r)\,dt^2+N^{-2}(r)\,dr^2+(N^{\phi}(r)\,dt+d\phi)^2.
\ee
where $N^{2}(r)$ and $N^{\phi}(r)$ are lapse and shift functions; 
\be
N^{2}(r)=-M+\frac{r^2}{l^2}+\frac{J^2}{4r^2},\qquad\qquad N^{\phi}(r)=-\frac{J}{2r^2},\nn
\ee
where $-\infty < t < \infty$ and $0\leq\phi\leq2\pi$. 
The solution is parameterized by the mass, $M$ and angular momentum $J$ of the black hole and they obey the conditions, $M>0$ and $|J|< Ml$. 

The proper distance from the horizon, $\rho$ is given by,
\be 
r^2=r_+^2\cosh^2\rho+ r_-^2\sinh^2\rho
\ee
where $r_+$ and $r_-$ are outer and inner horizons of the black hole. 

The metric of BTZ black hole can be written in term of proper distance \cite{DS1,DS2},
\be
ds^2=-\Big(u^2+\frac{J^2}{4l^2(u^2+M)}\,\Big)\,dt^2+d\rho^2+(\frac{J}{2l\sqrt{(u^2+M)}}dt-l\sqrt{{u^2+M}}d\phi)^2.
\ee
Where we are using $(r^2=l^2(u^2+M))$.

The Dirac equation in the background of BTZ black hole is given by (see appendix (\ref{A1})),
\ba
&&\Big[-\frac{1}{u}\tilde{\gamma}_0\partial_t+u\tilde{\gamma}_{1}\partial_{\rho}+\Big(\frac{-J}{2l^2u(u^2+M)}\tilde{\gamma}_0+\frac{1}{l\sqrt{u^2+M}}\tilde{\gamma}_2\Big)\partial_{\phi}\nn\\&&\qquad-\frac{1}{4}\Big(-2\tilde{\gamma}_1\Big(\frac{\sqrt{u^2+M}}{ul}-\frac{J^2}{4(u^2+M)^{3/2}}\Big)+\frac{J}{(u^2+M)\,l^2}\Big)-\mu I\Big]\Psi=0.\nn\\
\label{Dirac}
\ea
The two component Dirac spinor $\Psi$ can be written as;
\[\Psi = \left(\begin{array}{ccc}
\Psi_m^1(\rho)\Phi_m^{1} (\phi) \\
\Psi_m^2(\rho)\Phi_m^{2}(\phi) 
\end{array} \right)\]
where $\Phi_m^{1}=\frac{1}{\sqrt{2\pi}}\,e^{\iota\phi(m+1/2)}$ and $\Phi_m^{2}=\frac{1}{\sqrt{2\pi}}\,e^{\iota\phi(m-1/2)}$ are the eigen-vectors of the angular momentum operator.

Now, we can express the Hamiltonian of the system as the sum over azimuthal quantum number $m$, 
\be
H=\sum_{m}(\Psi^{\dagger 1}_{m},\Psi^{\dagger 2}_{m})\,H_m\left(\begin{array}{c} \Psi_{m}^1 \\
\Psi_{m}^2 
\end{array} \right).
\ee
The explicit form of $H_m$ can be obtained from (\ref{Dirac}) and is given by, 
\ba
H_m(\rho)&=&u^2\sigma_2\partial_{\rho}-\frac{u}{2}\Big(\frac{\sqrt{u^2+M}}{ul}-\frac{J^2}{4(u^2+M)^{3/2}}\Big)\sigma_2+\Big(\frac{-Jm}{2l^2(u^2+M)}\nn\\&&\qquad\qquad\qquad\qquad\qquad\qquad+\frac{um}{l\sqrt{(u^2+M)}}{\sigma}_1\Big)-\frac{u\,J}{(u^2+M)l^2}\sigma_3-\mu\, u\,\sigma_3,
\label{eqn:hamilton2}
\ea 
where we have also used $\tilde{\gamma}_0=\iota\sigma_3,\tilde{\gamma}_1=\sigma_1$, $\tilde{\gamma}_2=\sigma_2$, and $\sigma_3$ is multiplied by imaginary unit to change from Lorentzian to Euclidean signature.

We also set $\mu=0$ and $l=1$, in this and the next section. Massive case is considered in section (IV).

We discretize the system using the following relations \cite{Bombelli,Mukohyama:1998A,HM1},
\begin{align}
\rho\rightarrow(i-1/2)a,\qquad\qquad
\delta(\rho-\rho ')\rightarrow \delta_{ij}/a,\nn
\end{align}
where $i,~j=1,2 ....N$ and ``a'' is the lattice spacing . The continuum limit is obtained by taking $a\rightarrow 0$ and $N\rightarrow \infty$, while keeping the size of the system fixed. To discretize the Hamiltonian, we make the following replacements,
\ba
u[\rho=(i-\frac{1}{2})a]\rightarrow u_i, \qquad\qquad\Psi_m[\rho=(i-\frac{1}{2})a]\rightarrow\Psi^{i}_m, \nn
\ea
The discretized Hamiltonian of the system is given by (we also suppress angular momentum index 'm' here); \footnotetext[1]{We discretizing the Hamiltonian (\ref{eqn:hamilton2}) using the the middle-point prescription, the derivative of the form $f(x)\partial g(x)$ is replaced by $\frac{f_{j+1/2}[g_{j+1}−g_j]}{a}$ etc.}
\ba
H=&&\sum_{i,j}-\frac{i}{2}\Big(\Psi^{\dagger}_{i}u^2_{i+1/2}\sigma_2(\Psi_{i+1}-\Psi_{i})+\Psi^{\dagger}_{i}\frac{u_i}{2}\sigma_2\Big(\frac{\sqrt{u^2_i+M}}{u_i}-\frac{J^2}{4(u^2_i+M)^{3/2}}\Big)\Psi_{i}\nn\\&&+\Psi^{\dagger}_{i}\Big(\frac{-Jm}{2(u^2_i+M)}+\frac{m\,u_i}{l\sqrt{(u^2_i+M)}}{\sigma}_1\Big)\Psi_{i}-\Psi^{\dagger}_{i}\frac{J\,u_i}{(u^2_i+M)}\sigma_3\Psi_{i}\Big).
\label{discretehamil}
\ea 
The corresponding discrete Hamiltonian for an N lattice takes the form;
\be
H=\sum_{i,j=1}^N H_{i,j}=\sum_{i,j=1}^N(\psi^{\dagger}_{1 i},\psi^{\dagger}_{2i})\,M_{i,j}\left(\begin{array}{c} \psi_{1 j} \\
\psi_{2 j} 
\end{array} \right)\,,
\ee
where $i,j$ are the discrete variables corresponding to the radial distance $\rho$ and $M_{ij}$ for fixed $i,j$ is ($2 \times 2$) matrix such that,
\be
H_{ij}=\Psi^{\dagger}_{1i}\,M_{ij}^{11}\,\Psi_{1j}+\psi^{\dagger}_{1i}\,M_{ij}^{12}\,\Psi_{2j}+\Psi^{\dagger}_{2i}\,M_{ij}^{21}\,\Psi_{1j}+\Psi^{\dagger}_{2i}\,M_{ij}^{22}\,\Psi_{2j}.
\label{2.9}
\ee
For the general quadratic Hamiltonian, the fermion correlator appearing in the entanglement entropy formula is directly related to the $M_{i,j}$ by the relation (See appendix B),
\be 
C=\Theta(-M),
\label{cor}\ee
where $\Theta(x)$ is unit step function and $M$ is the trace of the matrix $M_{ij}$.

It is convenient to define ($2N \times 2N$) matrix ${\tilde M}^{2k+\alpha-2,2l+\beta-2}=M^{\alpha,\beta}_{k,l}$, for $ k,l=1,.....N$ and $\alpha,\beta=1,2$. The matrix elements of $\tilde{ M}_{kl} $ for fermion field in BTZ black hole case can be extracted from the Hamiltonian (\ref{discretehamil}) and are given explicitly as,

\ba
&&{\tilde M}^{kk}=(-1)^{k+1}\iota\Big[\frac{Jm}{{(u^2_i+M)}}+\frac{u_iJ}{{4(u^2_i+M)}}\Big],\nn\\
&&{\tilde M}^{2k-1,2k}=\iota\Big( u^2_{i+1/2}-\frac{u_i}{2}\Big(\frac{\sqrt{u^2_i+M}}{u_i}-\frac{J^2}{4(u^2_i+M)^{3/2}}\Big)\Big),\nn\\
&&{\tilde M}^{2k,2k-1}=-\iota\Big( u^2_{i+1/2}-\frac{u_i}{2}\Big(\frac{\sqrt{u^2_i+M}}{u_i}-\frac{J^2}{4(u^2_i+M)^{3/2}}\Big)\Big),\nn\\
&&{\tilde M}^{1,2}=\iota\Big( u^2_{i+1/2}-\frac{u_i}{2}\Big(\frac{\sqrt{u^2_i+M}}{u_i}-\frac{J^2}{4(u^2_i+M)^{3/2}}\Big)+\frac{m u_i}{\sqrt{u^2_i+M}}\Big),\nn\\
&& {\tilde M}^{2,1}=-\iota\Big( u^2_{i+1/2}-\frac{u_i}{2}\Big(\frac{\sqrt{u^2_i+M}}{u_i}-\frac{J^2}{4(u^2_i+M)^{3/2}}\Big)+\frac{m u_i}{\sqrt{u^2_i+M}}\Big),\nn\\
&&{\tilde M}^{2k-1,2k+2}=\frac{\iota}{2}\Big( \frac{m u_i}{\sqrt{u^2_i+M}}\Big),\qquad\qquad {\tilde M}^{2k,2k-3}=-\frac{\iota}{2}\Big( \frac{m u_i}{\sqrt{u^2_i+M}}\Big),\nn\\
&&{\tilde M}^{2k-1,2k-1}=\frac{\iota}{2}\Big( \frac{m u_i}{\sqrt{u^2_i+M}}\Big),\qquad\qquad \,\,{\tilde M}^{2k,2k+1}=-\frac{\iota}{2}\,\Big( \frac{m u_i}{\sqrt{u^2_i+M}}\Big).
\ea
These are the matrix elements of massless fields with angular momentum $J$. Since, the fermion correlator appearing in the entanglement entropy formula is related to $M_{ij}$, the matrix appearing in the Hamiltonian, for the general case the entropy of the system is given by a sum over the angular momentum ``$m$". We diagonalize the correlation matrix, and calculate the entanglement entropy, which can be expressed as, 
\be
S_{ent}= \sum_m-tr\Big[(1-C)\log(1-C)+C\log\,C\Big],
\label{eqn:entropy4}
\ee 
where $C$ is the correlation matrix related to the matrix $M$ by the equation (\ref{cor}). 
\section{Numerical Estimation of Entropy}
In order to calculate the entanglement entropy of the system, we split the total system $i=1,2...N$ into two subsystems labeled by $ (\alpha=1,2...n_B)$ and $(\beta=n_B+1,n_B+2...N)$ where $\alpha$ and $\beta$ are the inside and outside indices with respect to the position of spatial boundary (hypersurface) $R=n_B\,a$. The entropy is obtained by taking the limit; 
\be
S_{ent}=\lim_{N\rightarrow\infty}S(n_B,N),
\ee
where, $S(n_B,N)$ is the entanglement entropy of the total system $N$ with partition $n_B$. We consider the system discretized in radial direction with N=200 lattice points and partition size specified by the integer $n_B$. We obtain the reduced density matrix by tracing over degrees of freedom outside the hypersurface and calculate the entanglement entropy using equation (\ref{entferm}). The value of $n_B$ is taken in the range of $10 - 50$ and it suffices for extracting the term proportional to $R/a$ from $S(n_B,N )$. The numerical calculation of entropy begins with the calculation of matrix $\tilde M_{ij}$ and the two point correlator using the relation (\ref{cor}), and finally the entropy of the system (summed over angular momentum modes, m) is given by the equation (\ref{eqn:entropy4}). 

The partition size is given by $R=n_B a$ and the entropy scales with the size of the system and it can be a function of dimensionless ration $R/a$ in order to render the entropy finite. This argument also justifies the existence of a finite cut off length (here lattice spacing 'a') in the system. If we consider a hypersurface close to the black hole horizon (distance measured in units of $a$), the resulting entanglement entropy can be interpreted as black hole entropy and scales as $r_+/a$. We fit the numerical data in the following form,
\be
S_{ent}=C_s\Big(\frac{r_+}{a}\Big),
\ee 
where $C_s $ is the numerical constant, we estimate this coefficient numerically.

\begin{figure}[ht]
\centering
\includegraphics[width=.8\textwidth]{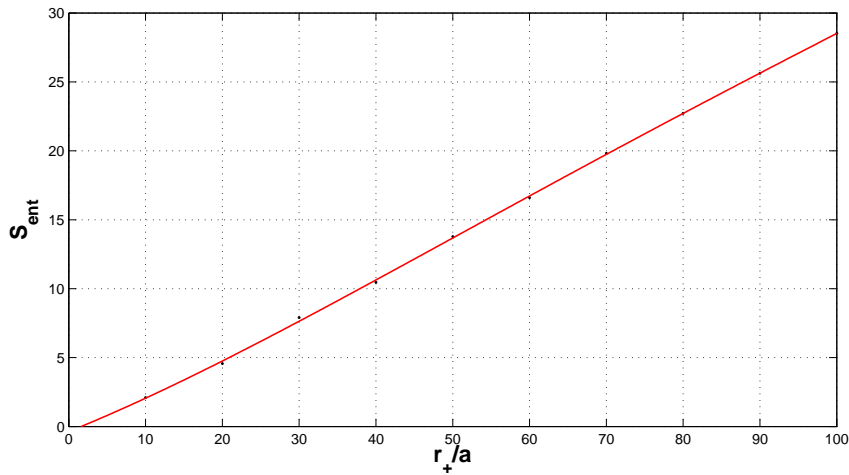}
\caption{The entanglement entropy of massless fermion field with the function of $r_+/a$.}
\label{fig:massless}
\end{figure}

We plot entropy as a function of $r_+/a$, where $r_+$ is the horizon of the black hole and this gives the value $c_s=.297$ (slope of line in figure (\ref{fig:massless})). 

The black hole entropy also receives quantum corrections given by logarithmic term in the entropy formula. The fitting procedure can also be used to calculate the co-efficient of logarithmic correction of entropy, 
\be
S_{log}=a\Big(\frac{r_+}{a}\Big)+b\Big(\log\frac{r_+}{a}\Big)+c.
\ee 

The numerical value of these coefficients is obtained by fitting procedure and is given as; a=0.304,~~b=-0.315,~~c=-0.327. 

The coefficient of logarithmic term for Dirac and scalar fields are related by the formula \cite{HMS, HC1},
\be
b_d^D=2^{[(d+1)/2]-1}b^s_d
\ee
Our numerical results seems to confirm this while using the results for the scalar field from our previous work \cite{DS1} and analytic result of Mann and Solodukhin \cite{MS}. 

\begin{figure}[ht]
\centering
\includegraphics[width=.8\textwidth]{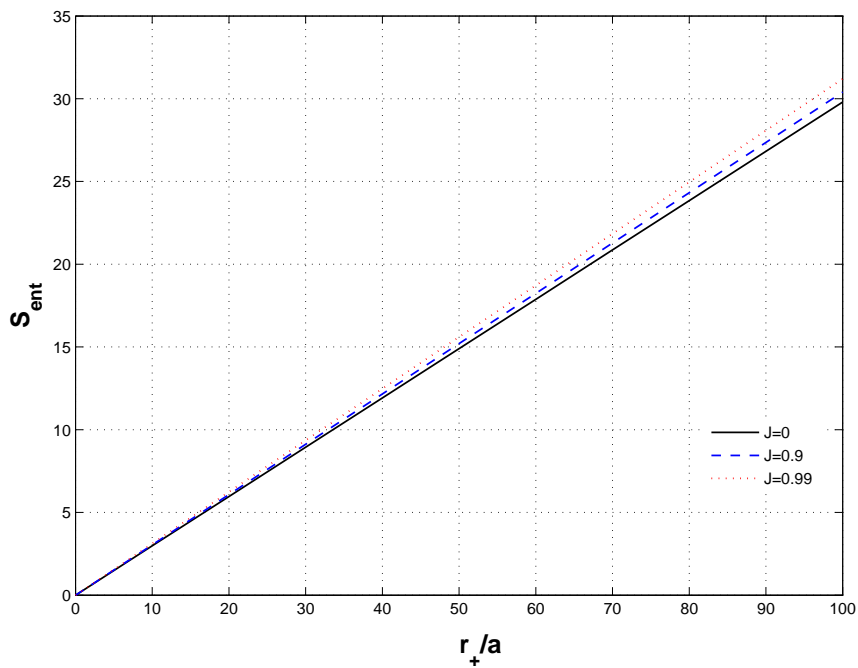}
\caption{The entanglement entropy of massless fermion field as a function of $r_+/a$ with different value of $J=0 ,0.9$ and $0.99$.}
\label{fig:angularmom}
\end{figure}

The dependence of entropy on angular momentum $J$ is also shown in the figure (\ref{fig:angularmom}). The entropy has no explicit dependence on angular momentum except through the factor $r_+/a$, which defines the size of the horizon. 
\newpage
\section{Entanglement Entropy in Free Massive theory}

In this section, we study the entanglement entropy in free massive theory in BTZ black hole space-time. The entropy can be expanded in powers of proper distance, $\rho$, for large values of $\rho$, \cite{H2},
\be
S=c_1(\mu)\rho+c_0(\mu)+c_{-1}(\mu)\frac{1}{\rho}+\ldots.
\ee
We calculate the entanglement entropy of the fermion field for different masses in the range $(.05<\mu<.5)$. 
The value of $c_0(\mu),c_1(\mu)$ and $c_{-1}(\mu)$ for different masses are tabulated in table (\ref{tab:mass}), 
\begin{center}
\begin{table}[h]
\begin{center}
\begin{tabular}{|l|l|r|l|r|l|r|}
\hline
\multicolumn{1}{|c|}{ } & \multicolumn{1}{c|}{$\mu=0.1$ } & \multicolumn{1}{c|}{ $\mu=0.2$ }& \multicolumn{1}{c|}{$\mu=0.3$} & \multicolumn{1}{c|}{$\mu=0.4$}& \multicolumn{1}{c|}{$\mu=0.5$}\\
\hline
\,\, $c_0(\mu)$\,\, & \,\, 0.23\,\, &0.21\,\, & \,\, 0.25& \,\, 0.32& \,\, 0.36\\
\,\, $c_1(\mu)$\,\, & \,\, 2.64\,\, &2.53\,\, & \,\, 2.42& \,\, 2.32& \,\, 2.22
\\
\,\, $c_{-1}(\mu)$\,\, & \,\, 1.20\,\, &0.84\,\, & \,\, 0.70& \,\, 0.64& \,\, 0.62
\\

\hline
\end{tabular}
\end{center}
\caption{The value of $c_0,c_1$ and $c_{-1}$ for different masses 0.1, 0.2, 0.3, 0.4 and 0.5.}
\label{tab:mass}
\end{table}
\end{center}
One can expand the $c_1(\mu)$ and $c_{-1}(\mu)$ in powers of $\mu$,
\be
c_1(\mu)=c_1\mu+c_1^0+c_{1}^{-1}\frac{1}{\mu},
\label{eqn:mass1}
\ee
\be
c_{-1}(\mu)=c_{-1}^1\mu+c_{-1}^0+c_{-1}\frac{1}{\mu}.
\label{eqn:mass2}
\ee

\begin{figure}
\centering
\includegraphics[width=.95\textwidth]{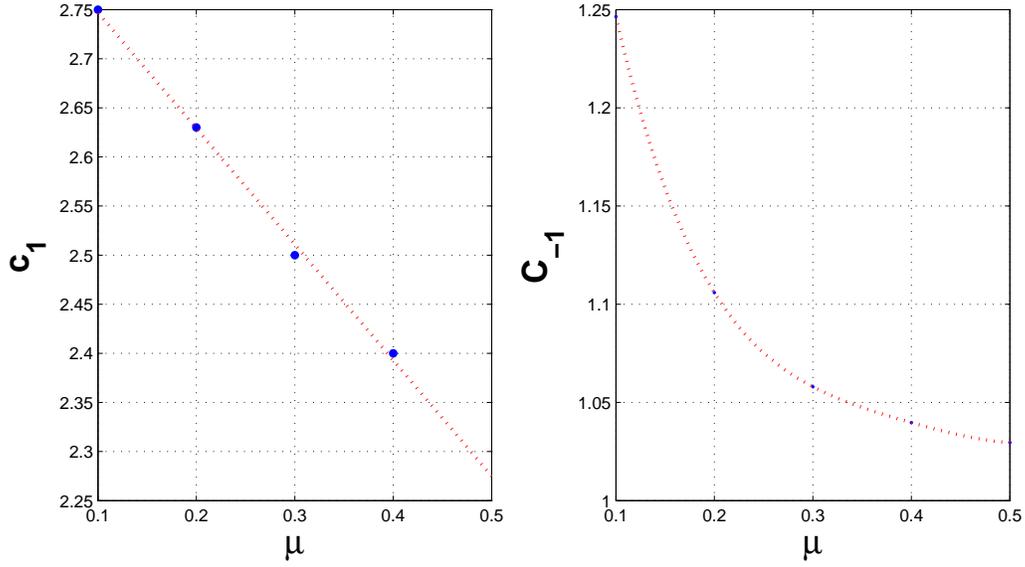}
\caption{The coefficients of $c_1$ and $c_{-1}$ are the function of $\mu$ with different masses. The co-efficient $c_1$ is proportional to $\mu$ , and $c_{-1}$ proportional to the $\frac{1}{\mu}$ in fitting.}
\label{fig:2}
\end{figure}

The plots of $c_1(\mu)$ and $c_{-1}(\mu)$ as a function of mass $\mu$ and $
1/\mu$ respectively for the fermion field are shown in figure (\ref{fig:2}). The co-efficients $c_1$ and $c_{-1}$ are found from the fitting the data plotted in figure (\ref{fig:2}) and the values are -0.503 and -0.074 respectively. 

\begin{figure}
\centering
\includegraphics[width=.8\textwidth]{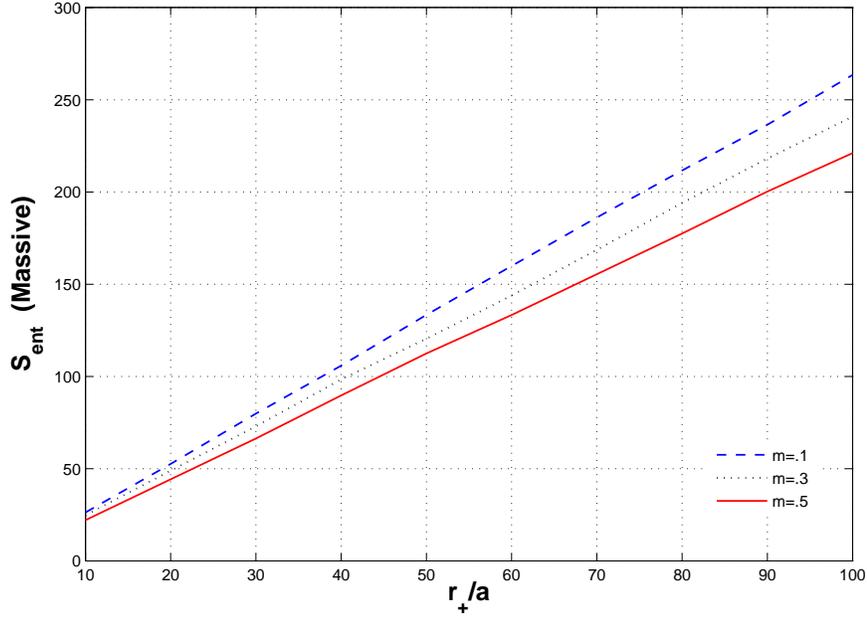}
\caption{The entanglement entropy of fermion field for massive fields of a function of $r_+/a$ with different masses (0.1, 0.3, and 0.5). } 
\label{figes}
\end{figure}

Here it is interesting to note that the massive theory in (2+1) dimension can be obtained via dimensional reduction of (3+1) dimensional massless theory \cite{Safdi,Safdi1}. The coefficients $c_{-1}$ is related with the co-efficient of logarithmic term in (3+1) dimension and is given by, $-\frac{\pi}{40}$ and $-\frac{\pi}{6}$.

\newpage
\section{Results and Conclusion}
In this paper, we have studied the entanglement entropy of the fermion field propagating in the background of BTZ black hole numerically. We have calculated the coefficients of leading and sub-leading correction of entropy for massless fermion field. The sub-leading correction gives the first quantum-correction of entropy, which is a logarithmic correction. We have calculated the coefficient of logarithmic term using fitting procedure. We also notice that the entropy of the fermion fields depends weakly upon the angular momentum. We have also studied the entanglement entropy for massive fermion fields numerically. 
One may also compute the entanglement entropy for fermion field numerically in the context of non-vacuum states (first excited state and mixed state) using the correlator method and study the correction of area relation. We leave it a future exercise.


\appendix
\section{Dirac Equation in BTZ Black Hole Background}
\label{A1}
The metric of BTZ Black Hole can be written in term of proper distance, $\rho$ defined as $r^2=r_+^2\cosh^2\rho+r_-^2\sinh^2\rho$, where $r_+$ and $r_-$ are outer and inner horizon of the black hole, 
\be
ds^2=-\Big(u^2+\frac{J^2}{4l^2(u^2+M)}\,\Big)\,dt^2+d\rho^2+(\frac{J}{2l\sqrt{(u^2+M)}}dt-l\sqrt{{u^2+M}}d\phi)^2,
\ee

and we use $\Big(r^2=l^2(u^2+M)\Big)$ etc. \cite{DS1}.

The Dirac equation in the curved space-time is written s;
\be
[\gamma^{\mu}\partial_{\mu}+\frac{1}{4}\gamma_{\mu}\omega_{\mu}^{ab}\gamma_{ab}-\mu I]\,\psi=0
\label{eq:gamma0}
\ee
where $\gamma^{\mu}$ are the Gamma matrices and $\omega^{ab}_{\mu}$ are spin-connection.

The Gamma matrices in the background of BTZ black hole space-time are,$\gamma_{\mu}=e^a_{\mu}\tilde{\gamma_a}$,
The {\it dreibein} are the given by;
\ba
&&e^0=\Big(u^2+\frac{J^2}{4l^2(u^2+M)}\,\Big)^{1/2}dt,\qquad dt=\Big(u^2+\frac{J^2}{4l^2(u^2+M)}\,\Big)^{-1/2}e^0\nn\\ 
&&e^1=d\rho,\qquad d\rho=e^1\nn\\
&&e^2=l\sqrt{u^2+M}\,d\phi-\frac{J}{2l\sqrt{(u^2+M)}}\,dt,\nn\\&&\qquad\qquad\qquad\qquad d\phi=\frac{1}{l\sqrt{(u^2+M)}}\,\Big[e^2+\sqrt{\Big(u^2+\frac{J^2}{4l^2(u^2+M)}\,\Big)}e^0\Big]
\ea

where
\begin{eqnarray}
&&\gamma_0=\sqrt{\Big(u^2+\frac{J^2}{4l^2(u^2+M)}\,\Big)}\,\tilde{\gamma}_0-\frac{J}{2l\sqrt{(u^2+M)}}\,\tilde{\gamma}_2,\qquad \nn\\
&&\gamma_1=\tilde{\gamma}_1,\qquad\qquad\qquad\qquad\qquad\gamma_2=l\sqrt{u^2+M}\,\tilde{\gamma}_2.
\end{eqnarray}
where $\tilde{\gamma}_a$ are the Pauli matrices in flat space-time and $\tilde{\gamma}_0=\iota\sigma_3,~~\tilde{\gamma}_1=\sigma_1,~~\tilde{\gamma}_2=\sigma_2.$ 
and the inverse of gamma matrices are, 
\ba
&&\gamma^0=\frac{1}{\sqrt{\Big(u^2+\frac{J^2}{4l^2(u^2+M)}\,\Big)}}\,\tilde{\gamma}_0,\qquad \gamma^1=\tilde{\gamma}^1\nn\\
&&\gamma^2=\frac{1}{l\sqrt{u^2+M}}\,\Big({\tilde{\gamma_2}}-\frac{J}{\sqrt{u^2+\frac{J^2}{4l^2(u^2+M)}}}\,{\tilde{\gamma_0}}\Big).
\ea

The non-vanishing Chritoffel symbols are, 
\ba
&&\Gamma_{10}^0=\Gamma_{01}^0= \frac{\sqrt{u^2+M}}{u^2+\frac{J^2}{4l^2{u^2+M}}},\qquad\qquad\qquad \Gamma^0_{21}=\Gamma^0_{12}=-\frac{J\sqrt{u^2+M}}{2(u^2+\frac{J^2}{4l^2({u^2+M)}})},\nn\\
&&\Gamma^1_{00}=\sqrt{u^2+M},\qquad\qquad\qquad\qquad\qquad\,\,
\Gamma^1_{22}=-l^2\,\sqrt{u^2+M},\nn\\
&&\Gamma^2_{10}=\Gamma^2_{01}=\frac{J\sqrt{u^2+M}}{2(u^2+\frac{J^2}{4l^2({u^2+M)}})},\qquad\qquad \Gamma^2_{21}=\Gamma^2_{12}=\frac{u^2}{\sqrt{u^2+M}\,(u^2+\frac{J^2}{4l^2({u^2+M)}})}.\nn\\
\ea

The non vanishing spin connections $\omega$ are,
\ba
&&\omega^0_{01}=\frac{1}{2}\Big[\frac{2}{l^2}-\frac{3}{2}\frac{J^2}{r^4}-\frac{2J^2}{r^3}\Big],\qquad\qquad \omega^0_{12}=\frac{J^2}{u(u^2+M)}\nn\\
&&\omega^0_{21}=\frac{J^2}{(u^2+M)},\qquad\qquad\qquad\qquad\,\,\,\omega^2_{01}=\frac{J^2u}{(u^2+M)}-\frac{Ju}{(u^2+M)},\nn\\
&&\omega^2_{10}=\frac{J^2}{u(u^2+M)},\qquad\qquad\qquad\qquad\omega^2_{21}=u.
\ea 

Finally, the Dirac equation (\ref{eq:gamma0}) in BTZ black hole space-time can be written as,

\ba
&&-\frac{1}{u}\tilde{\gamma}_0\partial_0+u\tilde{\gamma}_{1}\partial_1+\Big(\frac{-J}{2l^2u(u^2+M)}\tilde{\gamma}_0+\frac{1}{l\sqrt{u^2+M}}\tilde{\gamma}_2\Big)\partial_{2}\nn\\&&\qquad\qquad\qquad\qquad-\frac{1}{4}\Big(-2\tilde{\gamma}_1\Big(\frac{\sqrt{u^2+M}}{ul}-\frac{J^2}{4(u^2+M)^{3/2}}\Big)+\frac{J}{(u^2+M)\,l^2}\Big)-\mu I=0.\nn
\ea
where the subscript $(0,1,2)$ correspond to $(t,\rho,\phi)$.

\section{A Model of Entanglement Entropy for fermion fields}
In this section, we review the model of entanglement entropy for fermions and numerical computation of entropy. The appropriate formalism to describe the entanglement entropy is the density matrix, defined in the term of local operator $\cal O$ in the region $V$ of space and it is given by;
\be
tr(\rho_V {\cal O})=\langle0|{\cal O}|0\rangle.
\ee
The reduced density matrix of a system can be written in the exponential form \cite{HC1, HM1};
\be
\rho_V=ce^{-\cal H}
\label{1}
\ee
where ${\cal H}$ is the hermitian matrix to be identified with the Hamiltonian of the system and $c$ is the normalization constant.

The Hamiltonian of the system with fermions can be written as \cite{HM1}, 
\be
{\cal H}=\int_V \,\psi^{\dagger}_i\,H_{ij}\,\psi_j.
\label{2.21}
\ee
Using this form of the Hamiltonian the reduced density matrix $(\ref{1})$ is given by,
\be
\rho_V=ce^{\int_V \, \,\psi^{\dagger}_i\,H_{ij}\,\psi_j}.
\label{2}
\ee
We can diagonalize the exponent of density matrix by using the unitary transformation $d_l=U_{lm}\psi_m$, (one can choose $U$ such that, $UHU^{\dagger}=\{\epsilon_i\}$, $U$ being a unitary operator and $\epsilon_i$ are the eigenvalue of H).
The local creation and annihilation operator $\psi_i$ and $\psi_j^{\dagger}$ obey the anti commutation relations. The two point correlators are given as;
\ba
&&C_{ij}=\langle0|\,\psi_i\,\psi_j^{\dagger}|0\rangle,\qquad\qquad\langle0|\psi_j^{\dagger}\psi_i|0\rangle=\delta_{ij}-C_{ij},\nn\\
&&\langle0|\,\psi_i\,\psi_j|0\rangle=0,\qquad\quad\qquad\langle0|\psi_j^{\dagger}\psi_i^{\dagger}|0\rangle=0.
\label{2}
\ea 
The relation between $C$ and $H$ can be rewritten as ,
\be
C_{ij}=c\,tr\,\Big(e^{-\sum H_{lm}\psi_l^{\dagger}\psi_m}\psi_i\psi_j^{\dagger}\Big).
\label{6}
\ee 
For general quadratic case the discrete Hamiltonian is written as \cite{Huerta1}, 
\be
H=\sum_{i,j} {\psi}^\dagger_iM_{i,j}\,\, \psi_i
\ee
The correlator is directly related to $M_{i,j}$ by the relation,
\be
C=\Theta(-M)
\ee
where $M$ is the trace of the matrix and $\Theta$ is the step function. The entropy of the system is given by the relation,
\be
S=-tr[(1-C)\log(1-C)+C\log C].
\label{entferm}
\ee 
The eigenvalues of $C$ lies between $0$ and $1$, except in case where the global state is pure, then the eigenvalues of the $C$ are 0 or 1's. 

\newpage
\section*{Acknowledgments}
The work of Dharm Veer Singh is supported by Rajiv Gandhi National Fellowship Scheme, University Grant Commission (Under the fellowship award no. F.14-2(SC)/2008 (SA-III)) of Government of India.

\end{document}